\documentclass[aps,prd,reprint,superscriptaddress,showpacs,nofootinbib]{revtex4-1}

\usepackage{color}
\usepackage{amsmath}
\usepackage{amssymb}
\usepackage{graphicx}
\usepackage[normalem]{ulem}

\begin{document}
\title{Nuclear bound state of $\eta'(958)$ and partial restoration of
chiral symmetry in the $\eta^{\prime}$ mass}

\author{Daisuke Jido}
\affiliation{Yukawa Institute for Theoretical Physics, Kyoto University, 
Kyoto 606-8502, Japan}

\author{Hideko Nagahiro}
\affiliation{Department of Physics, Nara Women's University, 
Nara 630-8506, Japan}

\author{Satoru Hirenzaki}
\affiliation{Department of Physics, Nara Women's University, 
Nara 630-8506, Japan}


\date{\today}

\begin{abstract}
The in-medium mass of the $\eta^{\prime}$ meson is discussed 
in a context of partial restoration of chiral symmetry in nuclear medium. 
The $\eta^{\prime}$ mass is expected to be reduced by order of 100 MeV 
at the saturation density. The reduction is a consequence of the suppression of 
the anomaly effect on the $\eta^{\prime}$ mass induced by partial restoration 
of chiral symmetry. This strong attraction in $\eta^{\prime}$ nuclear systems 
does not accompany large absorption of $\eta^{\prime}$ into nuclear matter.
This leads to the possibility of so narrow bound states of the $\eta^{\prime}$ meson in nuclei to be seen in hadronic reactions with light nuclear targets.
\end{abstract}

\pacs{21.85.+d, 25.80.Hp, 14.40.Be}

\maketitle



The U$_{A}(1)$ problem~\cite{Weinberg:1975ui} has attracted continuous 
attention for a long time as a fundamental question on the low-energy 
spectrum and dynamics of the pseudoscalar mesons in QCD. 
Since quantum gluon dynamics explicitly breaks the U$_{A}(1)$ symmetry,
the $\eta^{\prime}$ meson is not necessarily a Nambu-Goldstone boson 
associated with spontaneous chiral symmetry breaking.
Thus, the peculiarly large mass of the $\eta^{\prime}$ meson
is a consequence of the quantum anomaly~\cite{Witten:1979vv} 
inducing the non-trivial vacuum structure of 
QCD~\cite{tHooft:1976fv,tHooft:1976up}.
%
It is also known that the $\eta^{\prime}$ spectrum strongly depends on 
the breaking pattern of chiral symmetry~\cite{Lee:1996zy}.

The study of dynamical chiral symmetry breaking and its partial 
restoration at finite density systems is one of the important subjects of 
contemporary hadron-nuclear physics. Recent experimental observations of 
pionic atoms~\cite{Friedman:2003wi}, especially deeply bound 
states in Sn isotopes~\cite{Suzuki:2002ae}, and 
low-energy pion-nucleus scattering~\cite{Friedman:2004jh,Friedman:2005pt} 
have figured out 
with helps of theoretical analyses~\cite{Kolomeitsev:2002gc,Jido:2008bk}
whether the partial restoration does take place in nuclei 
with order of 30\% reduction of the quark condensate. 
It goes a step further to the stage of precise determination of 
the density dependence of the quark condensate both in theory and 
experiment~\cite{Kaiser:2007nv,Ikeno:2011mv},
and systematic studies of the partial restoration appearing in other meson-nuclear systems.
As shown later, since the chiral symmetry breaking plays 
an important role also for the $\eta^{\prime}$ mass generation,
one expects strong mass reduction due to the partial restoration.

One of the 
efficient ways to observe in-medium modification of the meson properties 
is spectroscopy of meson-nucleus bound systems like deeply bound
pionic atoms. The main advantage
to observe the meson-nucleus bound 
system is that it is guaranteed that the meson inhabits the nucleus and 
it is unnecessary to remove in-vacuum contributions from the spectrum.
So far several meson nuclear bound states have been 
proposed~\cite{Haider:1986sa,Hayano:1998sy,Kishimoto:1999yj,Nagahiro:2004qz}
and experimental attempts of the bound state observation have been 
performed~\cite{Chrien:1988gn,Kishimoto:2003jr}. Nevertheless,
there are difficulties to observe clear signals for the mesonic bound 
states, because the bound states have 
large absorption widths due to strong interactions, such as 
conversion into lighter mesons and two nucleon absorptions~\cite{Kohno:1989wn,Yamagata:2006sm,Nagahiro:2008rj}.

The purposes of the present paper are twofold:
Firstly, we shed light upon the $\eta^{\prime}$ meson mass in nuclear 
matter in the context of partial restoration of chiral symmetry,
pointing out that the U$_{A}$(1) anomaly effects causes 
the $\eta^{\prime}$-$\eta$ mass difference necessarily through the chiral symmetry 
breaking. This fact leads to a relatively large mass reduction and 
weak absorption for $\eta^{\prime}$ in nuclear matter. 
Thus, we expect that possible nuclear bound states of 
the $\eta^{\prime}$ meson have narrower widths 
than their level spacings.
Secondly, we discuss experimental visibility of the $\eta^{\prime}$
bound states by showing typical formation spectra of the $\eta^{\prime}$ 
mesonic nuclei using a simple optical potential of $\eta^{\prime}$  in nuclei
which incorporates our theoretical consideration.

Theoretical investigations of the $\eta^{\prime}$ meson 
in finite energy density hadronic matter have been performed since
long ago~\cite{Pisarski:1983ms,Kunihiro:1989my,Kapusta:1995ww}, 
but the present status of the study of $\eta^{\prime}$-nucleus 
interaction is still exploratory due to lack of experimental information
and our knowledge of the fate of the $U_A(1)$ anomaly in nuclei
is rather short. 
The reduction of the $\eta^{\prime}$ mass in finite density systems
has been suggested in several theoretical approaches. For instance, 
{ 
it has been pointed out that 
the $\eta^{\prime}$ mass is reduced at finite density
due to rapid decrease of the instanton effects caused by strong 
suppression of the tunneling between different topological vacua~\cite{Kapusta:1995ww}.
}
In Nambu--Jona-Lasinio model calculations,
150 MeV mass reduction at the saturation density was 
suggested by using no density dependent determinant interaction~\cite{Costa:2002gk},
while with the density dependence 250 MeV reduction was reported 
in Ref.~\cite{Nagahiro:2006dr}.
Experimentally, 
it has been reported that a strong reduction of the $\eta^{\prime}$ 
mass, at least 200 MeV, is necessary to explain the two-pion correlation 
in Au + Au collisions at RHIC~\cite{PhysRevLett.105.182301,PhysRevC.83.054903}. 
On the other hand, 
analyses of the low-energy $\eta^{\prime}$ production experiment with $pp$ collisions
have suggested relatively smaller $\eta^{\prime}$-proton
scattering lengths, 
$|{\rm Re\ } a_{\eta^{\prime}p}| < 0.8$ fm~\cite{Moskal:2000gj}
and $|a_{\eta^{\prime}p}| \sim 0.1$ fm~\cite{Moskal:2000pu}, 
which correspond to from several to tens MeV mass reduction at the
nuclear saturation density if it is estimated by the linear density approximation. 

It is notable that 
the transparency ratios of the $\eta^{\prime}$ meson in nuclei were
also observed in TAPS and has suggested the absorption width of the $\eta^{\prime}$
meson at the saturation density is as small as around 30 MeV~\cite{Nanova:2011baryons}. 
In Ref.~\cite{Veneziano:1989ei}, it is reported that the 
$\eta^{\prime} NN$ three point
vertex should be suppressed according to an extended Goldberger-Treiman relation. 
Thus, the $\eta^{\prime}$ absorption into nuclear matter is possibly smaller. 

The basic idea of the present work is that one should distinguish 
between the anomaly operator itself and 
anomaly effects which are represented by matrix elements of the anomaly operator. 
The U$_{A}$(1) quantum anomaly appears
in the divergence of the flavor singlet axial vector current:
\begin{equation}
   \partial^{\mu} A_{\mu}^{(0)} = 2i (m_{u} \bar u \gamma_{5} u
   + m_{d} \bar d \gamma_{5} d + m_{s} \bar s \gamma_{5} s) 
   + \frac{3 \alpha_{s}}{8\pi} F \tilde F \label{eq:divA0} .
\end{equation}
The terms in the parentheses are the PCAC contributions and vanish in the chiral limit, 
while the last term in the right hand side is the U$_{A}$(1) anomaly term coming from 
gluon dynamics. Due to the last term the axial current does not conserve 
even in the chiral limit. Since Eq.~\eqref{eq:divA0} is an operator relation,
in order that the anomaly affects the $\eta^{\prime}$ mass, the operator $F\tilde F$
should couple to the $\eta^{\prime}$ state. This implies that 
it may happen that, even though the anomaly term is present in Eq.~\eqref{eq:divA0} 
and breaks the U$_{A}$(1) symmetry, 
the anomaly term does not couple to the $\eta^{\prime}$ state and 
the $\eta^{\prime}$ mass is not affected by the anomaly. 

We  see that this is the case when 
the SU(3)$_{L}\otimes$SU(3)$_{R}$ chiral symmetry is restored
in the following symmetry argument (see also a dynamical argument 
given in Ref.~\cite{Lee:1996zy}).
For simplicity we 
consider the three flavor chiral limit.
The mass spectra of the flavor singlet and octet pseudoscalar mesons is described  
by the correlation functions 
$\langle 0 | T \phi_{5}^{a}(x) \phi_{5}^{b\dagger}(0) | 0 \rangle$
with the pseudoscalar field $ \phi_{5}^{a} \equiv \bar q  i \gamma_{5} \lambda^{a} q$ 
($a=0,1,\dots,8$) with the quark field $q$ and the Gell-Mann matrix $\lambda^{a}$
for the SU(3) flavor. Because both flavor singlet and octet pseudoscalar fields
belong to the same $(\bf{3},\bf{\bar 3})\oplus (\bf{\bar 3},\bf{3})$ 
chiral multiplet of the SU(3)$_{L}\otimes$SU(3)$_{R}$ group,
when the SU(3)$_{L}\otimes$SU(3)$_{R}$ chiral symmetry is manifest,
the flavor singlet and octet spectra should degenerate,
no matter how the U$_{A}(1)$ anomaly depends on the density.
Therefore, this symmetry argument concludes that the $\eta$ and $\eta^{\prime}$ 
mass splitting can take place only with (dynamical and/or explicit) 
chiral symmetry breaking, meaning that  
the $U_{A}(1)$ anomaly effect does push the $\eta^{\prime}$ mass up 
necessarily with the chiral symmetry breaking.

In other words, the chiral singlet gluonic operator, which makes 
the $\eta^{\prime}$ mass lift up, cannot couple to the chiral pseudoscalar 
state without breaking chiral symmetry.
In the large-$N_{c}$ argument of  Refs.~\cite{Witten:1979vv,Veneziano:1979ec},
the $\eta^{\prime}$ mass can be obtained by the consistency 
condition that there should be cancellation between 
the flavor-singlet pseudoscalar pole and the gauge-dependent 
massless ghost in the correlation function of 
the topological charge density $F \tilde F$ in the soft limit. Performing 
the topological expansion of the quark loop, 
one has at the chiral limit~\cite{Witten:1979vv,Christos:1984tu}
\begin{equation}
   m_{\eta^{\prime}}^{2} = \frac{N_{c} c_{\eta^{\prime}}^{2}}{U_{0}(0)}
   \label{eq:MR}
\end{equation}
with the matrix element of $F\tilde F$
to create the $\eta^{\prime}$ meson
\begin{equation}
   \sqrt{N_{c}} c_{\eta^{\prime}} = \langle 0 | F \tilde F | \eta^{\prime} \rangle ,
\end{equation}
the number of color $N_{c}$ and $U_{0}(0)$ being the value of the
topological charge density correlator at the soft limit obtained 
without the quark loop, namely in the pure Yang-Mills theory.
Since chiral symmetry should be broken for the nonzero value 
of the matrix element $c_{\eta^{\prime}}$,
the mass relation (\ref{eq:MR}) shows that,
when chiral symmetry is being restored and the matrix element 
$c_{\eta^{\prime}}$ is getting reduced,
the mass of the flavor singlet $\eta^{\prime}$ 
should be going down, even if 
$F\tilde F$ appears in the divergence of the axial current.

In this way the mass splitting of the $\eta$-$\eta^{\prime}$ mesons is a 
consequence of the interplay of the U$_{A}(1)$ anomaly effect and the 
chiral symmetry breaking. 
Assuming 30\% reduction of the quark condensate in nuclear medium
and that the mass difference of $\eta$ and $\eta^{\prime}$ comes 
from the quark condensate linearly, one could expect an order of 150 MeV 
attraction for the $\eta^{\prime}$ meson coming from partial restoration 
of chiral symmetry in nuclear medium. 
This attraction is much stronger than, for instance, that for $\eta$ estimated in a chiral unitary model, which is as order of 50 MeV at the saturation density~\cite{Inoue:2002xw}.

The present mechanism of the $\eta^{\prime}$ mass reduction in finite 
density has another unique feature. 
In usual cases, attractive interactions of in-medium mesons induced by 
hadronic many-body effects unavoidably accompany comparably large 
absorptions. This can be perceived from the fact that the dispersion relation 
for the meson self-energy connects its real and imaginary parts as a consequence of 
the $s$-channel unitarity. This leads to the consequence that possible 
bound states have a comparable absorption width with the binding energy. 
For the attraction induced by gluon dynamics, like the present case, 
although some many-body effects introduce an absorptive potential
for the $\eta^{\prime}$ meson in medium,  
the mass reduction mechanism does not involve hadronic intermediate 
states and, thus, the attraction dose not accompany an additional imaginary part. 
Furthermore, in the present case, since the suppression of the U$_{A}(1)$ 
anomaly effect induces the attractive interaction, 
the influence acts selectively on the $\eta^{\prime}$ meson and, thus, 
it does not induce inelastic transitions of the $\eta^{\prime}$ meson into 
lighter mesons in nuclear medium.  
Consequently 
the $\eta^{\prime}$ absorption in nuclear matter can be small, which 
is consistent with the experimental finding~\cite{Nanova:2011baryons}.

As seen in the above observation, the $\eta^{\prime}$ mass is to be largely 
reduced in nuclear matter due to the suppression of the anomaly effect and 
simultaneously the absorption into nuclear matter can be small. 
Certainly with this attraction some bound states for $\eta^{\prime}$ in nuclei 
are formed. 
The question is whether the bound states are enough separated each other
with so narrow widths as to be observed in formation experiments. 
To observe clear signals of the bound states in formation experiments,
first of all,
it is important to choose appropriate nuclear targets of the reactions. 
Here we suggest nuclei with $A\sim 10$, such as 
$^{12}$C, as the target, since these nuclei may provide us with
finite density nuclear systems 
rather than systems with few-body characters. As
for heavier nuclei, we have several shell states 
for nucleons, which make the formation spectrum complicated
for the analyses as we will mention later.  

Let us sketch $\eta^{\prime}$ bound state structure in a nucleus
expected by the present argument and show formation spectra
of the $\eta^{\prime}$ mesonic nucleus in the  
$^{12}$C($\pi^{+},p)^{11}$C$\otimes\eta^{\prime}$ reaction.
We exploit a simple phenomenological optical potential of 
the $\eta^{\prime}$ meson in nuclei as
\begin{equation}
    V_{\eta^{\prime}}(r) = V_{0} \frac{\rho(r)}{\rho_{0}}, \label{eq:WSpot}
\end{equation}
with the Woods-Saxon type density distribution $\rho(r)$ for nucleus and 
the saturation density $\rho_0=0.17$ fm$^{-3}$. 
The depth of the attractive potential is a order of 100 MeV at the normal nuclear 
density as discussed above and the absorption width is
expected to be less than 40 MeV~\cite{Nanova:2011baryons} which
corresponds to the 20 MeV imaginary part of the optical potential.

\begin{figure}
   \includegraphics[width=\linewidth]{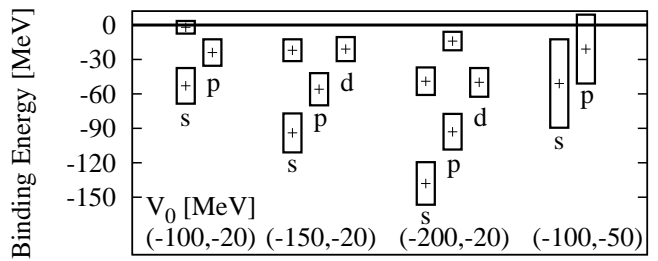}
\caption{Bound state spectra of the $\eta^{\prime}$ meson in
 $^{11}$C in units of MeV. Cross denotes the binding energy 
 and the band indicates the width of the bound state. 
The letters, s, p, d label the angular momentum states.
 The optical potential of the 
 $\eta^{\prime}$ meson in the nucleus is assumed in the form of
 Eq.~(\ref{eq:WSpot}) with the potential depths at the normal
 nuclear density 
 ${\rm Re}V_{0} = -100$, $-150$ and $-200$ MeV
 with a fixed imaginary potential ${\rm Im} V_{0} = 
 -20$ MeV.  We also show a result with the larger imaginary
 potential ${\rm Im}V_0=-50$ MeV with ${\rm Re}V_0=-100$ MeV. }
\label{fig:BS}
\end{figure}

In Fig.~\ref{fig:BS}, we show the bound state spectra
of the $\eta^{\prime}$ bound states, $B_{\eta} \equiv
E_{KG}-m_{\eta^{\prime}}$, in $^{11}$C, which is the 
residual nucleus in the $^{12}$C$(\pi^{+},p) $ reaction. 
The Klein-Gordon energy $E_{KG}$ is obtained as a complex value by solving 
the Klein-Gordon equation with the $\eta^{\prime}$ optical potential
in the form of Eq.~(\ref{eq:WSpot}) assuming the depths to be 
Re$V_{0}=-100$, $-150$, $-200$ MeV with Im$V_{0}=-20$
MeV. As seen in the figure, thanks to the strong attraction, there are 
several bound states in such a small nucleus, and, in addition, due 
to the small absorption these bound states are well separated. 
In contrast, as shown in Fig.~\ref{fig:BS}, 
optical potentials having a comparable imaginal part with the real part, such as
Re$V_{0}=-100$ MeV and Im$V_{0}=-50$ MeV,
provide bound states which have larger widths than the binding energies.
In this case it will be hard to observe these bound states as 
clear peaks in the formation spectra. 
For the detailed spectral structure of the bound states, 
the nuclear polarization will be important for the strongly 
interacting meson~\cite{Mares:2006vk}. 
We have checked, by assuming possible polarization of the core nucleus 
evaluated in a kaonic system with the 160 MeV binding energy~\cite{Mares:2006vk},  
that the core polarization effect shifts the bound state levels downwards but 
the bound states are so separated to be seen as isolated peaks.

In order to see the visibility of the peak structure of the
bound state spectrum in experiments, we calculate the formation 
spectra of the $\eta^{\prime}$ mesonic nuclei. 
We use the $^{12}$C$(\pi^{+},p)$ reaction
with the 1.8 GeV/c  incident $\pi^{+}$ beam to produce the 
$\eta^{\prime}$-nucleus system. Since the $\eta^{\prime}$ 
production in this reaction is exothermic, one cannot achieve 
the recoilless condition for the $\eta^{\prime}$ in nuclei. 
One observes the spectrum of the $\eta^{\prime}$-nucleus system 
by detecting the emitted proton at the forward
direction in order to reduce the momentum transfer.

The formation spectrum is calculated in the approach developed  
in Refs.~\cite{Jido:2002yb,Nagahiro:2003iv,Nagahiro:2005gf,Jido:2008ng,Nagahiro:2008rj}.
The calculated spectrum is scaled by the forward differential cross section of
the elementary $\pi^+n\rightarrow\eta' p$ process, which is 
estimated to be 100 $\mu$b/sr in the laboratory frame
from the total cross section $\sigma\sim 100$ $\mu$b~\cite{Rader:1973mx} 
under the assumption of isotropic angular dependence in the center of mass frame.
We calculate the formation spectra separately in the subcomponents of
the $\eta'$-mesonic nuclei labeled by
$(n\ell_j)_n^{-1}\otimes\ell_{\eta'}$ that indicates the formation of 
an $\eta'$ meson in the $\ell_{\eta'}$ orbit with
a neutron-hole in the $\ell$ orbit with the total spin $j$ and the
principal quantum number $n$ in the daughter nucleus.
The calculated spectra are shown as functions of 
$E_{\rm ex}-E_{0}$ where $E_{0}$ is the $\eta^{\prime}$ production threshold 
with the ground state daughter nucleus
and the excitation energy $E_{\rm ex}$ is defined by 
$E_{\rm ex} \equiv m_{\eta'} -B_{\eta'} + [S_n(j_n)-S_n(0p_{3/2})]$ 
with $S_n(j_n)$ being the neutron separation energy from the neutron single-particle 
level $j_n$ to take into account the difference of the separation energy 
$S_n(j_n)-S_n(0p_{3/2})=18$~MeV for the subcomponents accompanied 
by the $(0s_{1/2})_n^{-1}$ hole-state.  

\begin{figure*}
   \includegraphics[width=\linewidth]{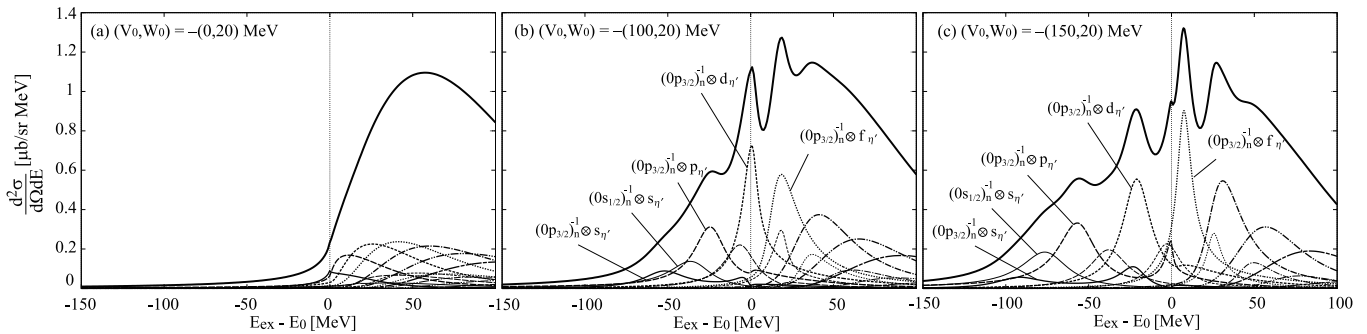}
\caption{{Calculated spectra of the
 $^{12}$C($\pi^+,p)^{11}$C$\otimes\eta'$ at $p_\pi=1.8$ GeV/c as functions
 of the exitation energy $E_{\rm ex}$ with (a) $V_0=-(0+20i)$ MeV, (b)
 $V_0=-(100+20i)$ MeV and (c) $V_0=-(150+20i)$ MeV.  The thick solid lines
 show the total spectra, and the dominant subcomponents are labeled
 by the neutron-hole state $(n\ell_j)_n^{-1}$ and the $\eta'$ state $\ell_{\eta'}$.
}}
\label{fig:spec}
\end{figure*}

In Fig.~\ref{fig:spec}, we show the
calculated $^{12}$C$(\pi^+,p)^{11}$C$\otimes\eta'$ cross
sections with three different potential parameters. (See the figure caption.)
In the figure, the vertical line at $E_{\rm ex}-E_{0}=0$ indicates 
the $\eta^{\prime}$ production threshold in vacuum. 
In the case of no attractive potential, there is no structure in the 
$\eta^{\prime}$-binding  region but some bump in the quasi-free region. 
Taking this case as a 
reference, we discuss the structure of the formation spectra with 
the attractive potentials. 
In each plot, the subcomponents with the $(0s_{1/2})_n^{-1}$ hole-state
give less contributions, because there are only half the neutrons in the 
$s_{1/2}$ state of the $p_{3/2}$ neutrons in the parent nucleus. 
Finding so prominent peaks in the $\eta^{\prime}$-binding region
as to be possibly observed in future experiments, we conclude that 
with an order of 100 MeV mass reduction and a 40 MeV absorption width 
at the saturation density we have a chance to observe 
the $\eta^{\prime}$-nucleus bound states in the $^{12}$C$(\pi^{+},p)$ reaction.
We see also clear peaks around the $\eta^{\prime}$ production threshold,
for instance $(0p_{3/2})_{n}^{-1}\otimes d_{\eta^{\prime}}$ in plot (b)
and $(0p_{3/2})_{n}^{-1}\otimes f_{\eta^{\prime}}$ in plot (c). They are 
not signals of the bound states, because no bound states exist
in the $d$ and $f$ states for the case (b) and (c), respectively,
as shown in Fig.~\ref{fig:BS}. Nevertheless, these are 
remnants of the bound states which could be formed if the attraction 
would be stronger. Therefore, such peak structure also can be 
signals of the strong attractive potential.

Similarly to hypernucleus production spectra in the $(\pi^{+},K^{+})$
reaction, in the obtained spectra shown in Fig.~\ref{fig:spec}
many subcomponents with different quantum numbers give certain 
contributions because of the finite momentum transfer (200 MeV/c) in 
the present reaction.  Thus, to identify the quantum number of each peak,
precise measurements and theoretical analyses are
necessary. Nevertheless, observing peak structure is the important 
first step to perform detailed spectroscopy of the $\eta^{\prime}$ bound states.
As seen in Refs.~\cite{Nagahiro:2004qz,Nagahiro:2006dr,Nagahiro:2010zz},
the structure of the formation spectra is not so dependent on the 
formation reaction of the $\eta^{\prime}$ mesonic nuclei.

The experimental feasibility for the observation of the peak structure 
highly depends on the level spacing of the bound states 
and their widths. Since $\eta^{\prime}$ is in the Wood-Saxon type
potential induced by the nuclear density, the level spacing of the 
$\eta^{\prime}$ bound states in a nucleus with mass number $A$
is characterized as $\hbar \omega \sim 40 A^{-1/3}$ MeV similarliy to 
the major shell spacing for nuclei. 
For observation of clear peak structure, the level spacing should be
larger than the level width $\Gamma\sim -2 {\rm Im}V$.
For the $^{12}$C target case, the upper limit would be 
$\Gamma = -2 {\rm Im}V \sim 50$ MeV, which is larger than 
$\Gamma=25$--$30$ MeV extracted from the transparency 
ratio at $p=0.95$ GeV/c in Ref.~\cite{Nanova:2011baryons}.
For larger $A$, even though the formation cross section can be larger, 
the peak structure gets less prominent because of the following reason.  
The bound state spectrum is determined by convolutions
of the nucleon hole and $\eta^{\prime}$ bound wavefunction. 
For larger $A$, there are more levels of the hole state and the level spacing 
of the $\eta^{\prime}$ bound states are smaller. Consequently the 
peaks coming from many possible combinations get overlapped and 
the peak structure is smeared out. Thus, nuclei with $A\sim$ 10 to 20 are
good candidates of the target for the formation experiments.

In conclusion, 
we point out that partial restoration of chiral symmetry in a nuclear medium 
induces suppression of the U$_{A}(1)$ anomaly {\it effect} to the $\eta^{\prime}$ mass.
Consequently, we expect a large mass reduction of the $\eta^{\prime}$ meson 
in nuclear matter with relatively smaller absorption. The mass reduction 
could be observed as $\eta^{\prime}$-nucleus bound states in the formation reactions. 
The interplay between the chiral symmetry restoration 
and the U$_{A}(1)$ anomaly effect can be a clue 
to understand the $\eta^{\prime}$ mass generation mechanism. Therefore,
experimental observations of the deeply $\eta^{\prime}$-nucleus bound states, or 
even confirmation of nonexistence of such deeply bound states,
is important to understand the U$_{A}(1)$ anomaly effects on hadrons.

{\it Acknowledgement.} The authors would like to thank Dr.\ K.\ Itahashi, Dr.\ H.\ Fujioka and Prof.\ W.\ Weise 
for useful discussion.
This work was partially supported by the Grants-in-Aid for Scientific Research (No. 22740161, No. 20540273, and No. 22105510). This work was done in part under the Yukawa International Program for Quark- hadron Sciences (YIPQS).


\begin{thebibliography}{43}%
\makeatletter
\providecommand \@ifxundefined [1]{%
 \@ifx{#1\undefined}
}%
\providecommand \@ifnum [1]{%
 \ifnum #1\expandafter \@firstoftwo
 \else \expandafter \@secondoftwo
 \fi
}%
\providecommand \@ifx [1]{%
 \ifx #1\expandafter \@firstoftwo
 \else \expandafter \@secondoftwo
 \fi
}%
\providecommand \natexlab [1]{#1}%
\providecommand \enquote  [1]{``#1''}%
\providecommand \bibnamefont  [1]{#1}%
\providecommand \bibfnamefont [1]{#1}%
\providecommand \citenamefont [1]{#1}%
\providecommand \href@noop [0]{\@secondoftwo}%
\providecommand \href [0]{\begingroup \@sanitize@url \@href}%
\providecommand \@href[1]{\@@startlink{#1}\@@href}%
\providecommand \@@href[1]{\endgroup#1\@@endlink}%
\providecommand \@sanitize@url [0]{\catcode `\\12\catcode `\$12\catcode
  `\&12\catcode `\#12\catcode `\^12\catcode `\_12\catcode `\%12\relax}%
\providecommand \@@startlink[1]{}%
\providecommand \@@endlink[0]{}%
\providecommand \url  [0]{\begingroup\@sanitize@url \@url }%
\providecommand \@url [1]{\endgroup\@href {#1}{\urlprefix }}%
\providecommand \urlprefix  [0]{URL }%
\providecommand \Eprint [0]{\href }%
\providecommand \doibase [0]{http://dx.doi.org/}%
\providecommand \selectlanguage [0]{\@gobble}%
\providecommand \bibinfo  [0]{\@secondoftwo}%
\providecommand \bibfield  [0]{\@secondoftwo}%
\providecommand \translation [1]{[#1]}%
\providecommand \BibitemOpen [0]{}%
\providecommand \bibitemStop [0]{}%
\providecommand \bibitemNoStop [0]{.\EOS\space}%
\providecommand \EOS [0]{\spacefactor3000\relax}%
\providecommand \BibitemShut  [1]{\csname bibitem#1\endcsname}%
\let\auto@bib@innerbib\@empty
\bibitem [{\citenamefont {Weinberg}(1975)}]{Weinberg:1975ui}%
  \BibitemOpen
  \bibfield  {author} {\bibinfo {author} {\bibfnamefont {S.}~\bibnamefont
  {Weinberg}},\ }\href {\doibase 10.1103/PhysRevD.11.3583} {\bibfield
  {journal} {\bibinfo  {journal} {Phys. Rev.}\ }\textbf {\bibinfo {volume}
  {D11}},\ \bibinfo {pages} {3583} (\bibinfo {year} {1975})}\BibitemShut
  {NoStop}%
\bibitem [{\citenamefont {Witten}(1979)}]{Witten:1979vv}%
  \BibitemOpen
  \bibfield  {author} {\bibinfo {author} {\bibfnamefont {E.}~\bibnamefont
  {Witten}},\ }\href {\doibase 10.1016/0550-3213(79)90031-2} {\bibfield
  {journal} {\bibinfo  {journal} {Nucl. Phys.}\ }\textbf {\bibinfo {volume}
  {B156}},\ \bibinfo {pages} {269} (\bibinfo {year} {1979})}\BibitemShut
  {NoStop}%
\bibitem [{\citenamefont {'t~Hooft}(1976{\natexlab{a}})}]{tHooft:1976fv}%
  \BibitemOpen
  \bibfield  {author} {\bibinfo {author} {\bibfnamefont {G.}~\bibnamefont
  {'t~Hooft}},\ }\href {\doibase 10.1103/PhysRevD.14.3432} {\bibfield
  {journal} {\bibinfo  {journal} {Phys. Rev.}\ }\textbf {\bibinfo {volume}
  {D14}},\ \bibinfo {pages} {3432} (\bibinfo {year}
  {1976}{\natexlab{a}})}\BibitemShut {NoStop}%
\bibitem [{\citenamefont {'t~Hooft}(1976{\natexlab{b}})}]{tHooft:1976up}%
  \BibitemOpen
  \bibfield  {author} {\bibinfo {author} {\bibfnamefont {G.}~\bibnamefont
  {'t~Hooft}},\ }\href {\doibase 10.1103/PhysRevLett.37.8} {\bibfield
  {journal} {\bibinfo  {journal} {Phys. Rev. Lett.}\ }\textbf {\bibinfo
  {volume} {37}},\ \bibinfo {pages} {8} (\bibinfo {year}
  {1976}{\natexlab{b}})}\BibitemShut {NoStop}%
\bibitem [{\citenamefont {Lee}\ and\ \citenamefont
  {Hatsuda}(1996)}]{Lee:1996zy}%
  \BibitemOpen
  \bibfield  {author} {\bibinfo {author} {\bibfnamefont {S.~H.}\ \bibnamefont
  {Lee}}\ and\ \bibinfo {author} {\bibfnamefont {T.}~\bibnamefont {Hatsuda}},\
  }\href {\doibase 10.1103/PhysRevD.54.R1871} {\bibfield  {journal} {\bibinfo
  {journal} {Phys. Rev.}\ }\textbf {\bibinfo {volume} {D54}},\ \bibinfo {pages}
  {1871} (\bibinfo {year} {1996})}\BibitemShut {NoStop}%
\bibitem [{\citenamefont {Friedman}\ and\ \citenamefont
  {Gal}(2003)}]{Friedman:2003wi}%
  \BibitemOpen
  \bibfield  {author} {\bibinfo {author} {\bibfnamefont {E.}~\bibnamefont
  {Friedman}}\ and\ \bibinfo {author} {\bibfnamefont {A.}~\bibnamefont {Gal}},\
  }\href {\doibase 10.1016/S0375-9474(03)01476-3} {\bibfield  {journal}
  {\bibinfo  {journal} {Nucl. Phys.}\ }\textbf {\bibinfo {volume} {A724}},\
  \bibinfo {pages} {143} (\bibinfo {year} {2003})}\BibitemShut {NoStop}%
\bibitem [{\citenamefont {Suzuki}\ \emph {et~al.}(2004)\citenamefont {Suzuki}
  \emph {et~al.}}]{Suzuki:2002ae}%
  \BibitemOpen
  \bibfield  {author} {\bibinfo {author} {\bibfnamefont {K.}~\bibnamefont
  {Suzuki}} \emph {et~al.},\ }\href {\doibase 10.1103/PhysRevLett.92.072302}
  {\bibfield  {journal} {\bibinfo  {journal} {Phys. Rev. Lett.}\ }\textbf
  {\bibinfo {volume} {92}},\ \bibinfo {pages} {072302} (\bibinfo {year}
  {2004})}\BibitemShut {NoStop}%
\bibitem [{\citenamefont {Friedman}\ \emph {et~al.}(2004)\citenamefont
  {Friedman} \emph {et~al.}}]{Friedman:2004jh}%
  \BibitemOpen
  \bibfield  {author} {\bibinfo {author} {\bibfnamefont {E.}~\bibnamefont
  {Friedman}} \emph {et~al.},\ }\href {\doibase 10.1103/PhysRevLett.93.122302}
  {\bibfield  {journal} {\bibinfo  {journal} {Phys. Rev. Lett.}\ }\textbf
  {\bibinfo {volume} {93}},\ \bibinfo {pages} {122302} (\bibinfo {year}
  {2004})}\BibitemShut {NoStop}%
\bibitem [{\citenamefont {Friedman}\ \emph {et~al.}(2005)\citenamefont
  {Friedman} \emph {et~al.}}]{Friedman:2005pt}%
  \BibitemOpen
  \bibfield  {author} {\bibinfo {author} {\bibfnamefont {E.}~\bibnamefont
  {Friedman}} \emph {et~al.},\ }\href {\doibase 10.1103/PhysRevC.72.034609}
  {\bibfield  {journal} {\bibinfo  {journal} {Phys. Rev.}\ }\textbf {\bibinfo
  {volume} {C72}},\ \bibinfo {pages} {034609} (\bibinfo {year}
  {2005})}\BibitemShut {NoStop}%
\bibitem [{\citenamefont {Kolomeitsev}\ \emph {et~al.}(2003)\citenamefont
  {Kolomeitsev}, \citenamefont {Kaiser},\ and\ \citenamefont
  {Weise}}]{Kolomeitsev:2002gc}%
  \BibitemOpen
  \bibfield  {author} {\bibinfo {author} {\bibfnamefont {E.~E.}\ \bibnamefont
  {Kolomeitsev}}, \bibinfo {author} {\bibfnamefont {N.}~\bibnamefont {Kaiser}},
  \ and\ \bibinfo {author} {\bibfnamefont {W.}~\bibnamefont {Weise}},\
  }\href@noop {} {\bibfield  {journal} {\bibinfo  {journal} {Phys. Rev. Lett.}\
  }\textbf {\bibinfo {volume} {90}},\ \bibinfo {pages} {092501} (\bibinfo
  {year} {2003})}\BibitemShut {NoStop}%
\bibitem [{\citenamefont {Jido}\ \emph
  {et~al.}(2008{\natexlab{a}})\citenamefont {Jido}, \citenamefont {Hatsuda},\
  and\ \citenamefont {Kunihiro}}]{Jido:2008bk}%
  \BibitemOpen
  \bibfield  {author} {\bibinfo {author} {\bibfnamefont {D.}~\bibnamefont
  {Jido}}, \bibinfo {author} {\bibfnamefont {T.}~\bibnamefont {Hatsuda}}, \
  and\ \bibinfo {author} {\bibfnamefont {T.}~\bibnamefont {Kunihiro}},\ }\href
  {\doibase 10.1016/j.physletb.2008.10.034} {\bibfield  {journal} {\bibinfo
  {journal} {Phys. Lett.}\ }\textbf {\bibinfo {volume} {B670}},\ \bibinfo
  {pages} {109} (\bibinfo {year} {2008}{\natexlab{a}})}\BibitemShut {NoStop}%
\bibitem [{\citenamefont {Kaiser}\ \emph {et~al.}(2008)\citenamefont {Kaiser},
  \citenamefont {de~Homont},\ and\ \citenamefont {Weise}}]{Kaiser:2007nv}%
  \BibitemOpen
  \bibfield  {author} {\bibinfo {author} {\bibfnamefont {N.}~\bibnamefont
  {Kaiser}}, \bibinfo {author} {\bibfnamefont {P.}~\bibnamefont {de~Homont}}, \
  and\ \bibinfo {author} {\bibfnamefont {W.}~\bibnamefont {Weise}},\ }\href
  {\doibase 10.1103/PhysRevC.77.025204} {\bibfield  {journal} {\bibinfo
  {journal} {Phys. Rev.}\ }\textbf {\bibinfo {volume} {C77}},\ \bibinfo {pages}
  {025204} (\bibinfo {year} {2008})}\BibitemShut {NoStop}%
\bibitem [{\citenamefont {Ikeno}\ \emph {et~al.}(2011)\citenamefont {Ikeno}
  \emph {et~al.}}]{Ikeno:2011mv}%
  \BibitemOpen
  \bibfield  {author} {\bibinfo {author} {\bibfnamefont {N.}~\bibnamefont
  {Ikeno}} \emph {et~al.},\ }\href {\doibase 10.1143/PTP.126.483} {\bibfield
  {journal} {\bibinfo  {journal} {Prog. Theor. Phys.}\ }\textbf {\bibinfo
  {volume} {126}},\ \bibinfo {pages} {483} (\bibinfo {year}
  {2011})}\BibitemShut {NoStop}%
\bibitem [{\citenamefont {Haider}\ and\ \citenamefont
  {Liu}(1986)}]{Haider:1986sa}%
  \BibitemOpen
  \bibfield  {author} {\bibinfo {author} {\bibfnamefont {Q.}~\bibnamefont
  {Haider}}\ and\ \bibinfo {author} {\bibfnamefont {L.~C.}\ \bibnamefont
  {Liu}},\ }\href {\doibase 10.1016/0370-2693(86)90846-4} {\bibfield  {journal}
  {\bibinfo  {journal} {Phys. Lett.}\ }\textbf {\bibinfo {volume} {B172}},\
  \bibinfo {pages} {257} (\bibinfo {year} {1986})}\BibitemShut {NoStop}%
\bibitem [{\citenamefont {Hayano}\ \emph {et~al.}(1999)\citenamefont {Hayano},
  \citenamefont {Hirenzaki},\ and\ \citenamefont {Gillitzer}}]{Hayano:1998sy}%
  \BibitemOpen
  \bibfield  {author} {\bibinfo {author} {\bibfnamefont {R.~S.}\ \bibnamefont
  {Hayano}}, \bibinfo {author} {\bibfnamefont {S.}~\bibnamefont {Hirenzaki}}, \
  and\ \bibinfo {author} {\bibfnamefont {A.}~\bibnamefont {Gillitzer}},\ }\href
  {\doibase 10.1007/s100500050322} {\bibfield  {journal} {\bibinfo  {journal}
  {Eur. Phys. J.}\ }\textbf {\bibinfo {volume} {A6}},\ \bibinfo {pages} {99}
  (\bibinfo {year} {1999})}\BibitemShut {NoStop}%
\bibitem [{\citenamefont {Kishimoto}(1999)}]{Kishimoto:1999yj}%
  \BibitemOpen
  \bibfield  {author} {\bibinfo {author} {\bibfnamefont {T.}~\bibnamefont
  {Kishimoto}},\ }\href {\doibase 10.1103/PhysRevLett.83.4701} {\bibfield
  {journal} {\bibinfo  {journal} {Phys. Rev. Lett.}\ }\textbf {\bibinfo
  {volume} {83}},\ \bibinfo {pages} {4701} (\bibinfo {year}
  {1999})}\BibitemShut {NoStop}%
\bibitem [{\citenamefont {Nagahiro}\ and\ \citenamefont
  {Hirenzaki}(2005)}]{Nagahiro:2004qz}%
  \BibitemOpen
  \bibfield  {author} {\bibinfo {author} {\bibfnamefont {H.}~\bibnamefont
  {Nagahiro}}\ and\ \bibinfo {author} {\bibfnamefont {S.}~\bibnamefont
  {Hirenzaki}},\ }\href {\doibase 10.1103/PhysRevLett.94.232503} {\bibfield
  {journal} {\bibinfo  {journal} {Phys. Rev. Lett.}\ }\textbf {\bibinfo
  {volume} {94}},\ \bibinfo {pages} {232503} (\bibinfo {year}
  {2005})}\BibitemShut {NoStop}%
\bibitem [{\citenamefont {Chrien}\ \emph {et~al.}(1988)\citenamefont {Chrien}
  \emph {et~al.}}]{Chrien:1988gn}%
  \BibitemOpen
  \bibfield  {author} {\bibinfo {author} {\bibfnamefont {R.~E.}\ \bibnamefont
  {Chrien}} \emph {et~al.},\ }\href@noop {} {\bibfield  {journal} {\bibinfo
  {journal} {Phys. Rev. Lett.}\ }\textbf {\bibinfo {volume} {60}},\ \bibinfo
  {pages} {2595} (\bibinfo {year} {1988})}\BibitemShut {NoStop}%
\bibitem [{\citenamefont {Kishimoto}\ \emph {et~al.}(2003)\citenamefont
  {Kishimoto} \emph {et~al.}}]{Kishimoto:2003jr}%
  \BibitemOpen
  \bibfield  {author} {\bibinfo {author} {\bibfnamefont {T.}~\bibnamefont
  {Kishimoto}} \emph {et~al.},\ }\href@noop {} {\bibfield  {journal} {\bibinfo
  {journal} {Prog. Theor. Phys. Suppl.}\ }\textbf {\bibinfo {volume} {149}},\
  \bibinfo {pages} {264} (\bibinfo {year} {2003})}\BibitemShut {NoStop}%
\bibitem [{\citenamefont {Kohno}\ and\ \citenamefont
  {Tanabe}(1989)}]{Kohno:1989wn}%
  \BibitemOpen
  \bibfield  {author} {\bibinfo {author} {\bibfnamefont {M.}~\bibnamefont
  {Kohno}}\ and\ \bibinfo {author} {\bibfnamefont {H.}~\bibnamefont {Tanabe}},\
  }\href {\doibase 10.1016/0370-2693(89)90202-5} {\bibfield  {journal}
  {\bibinfo  {journal} {Phys. Lett.}\ }\textbf {\bibinfo {volume} {B231}},\
  \bibinfo {pages} {219} (\bibinfo {year} {1989})}\BibitemShut {NoStop}%
\bibitem [{\citenamefont {Yamagata}\ \emph {et~al.}(2006)\citenamefont
  {Yamagata}, \citenamefont {Nagahiro},\ and\ \citenamefont
  {Hirenzaki}}]{Yamagata:2006sm}%
  \BibitemOpen
  \bibfield  {author} {\bibinfo {author} {\bibfnamefont {J.}~\bibnamefont
  {Yamagata}}, \bibinfo {author} {\bibfnamefont {H.}~\bibnamefont {Nagahiro}},
  \ and\ \bibinfo {author} {\bibfnamefont {S.}~\bibnamefont {Hirenzaki}},\
  }\href {\doibase 10.1103/PhysRevC.74.014604} {\bibfield  {journal} {\bibinfo
  {journal} {Phys. Rev.}\ }\textbf {\bibinfo {volume} {C76}},\ \bibinfo {pages}
  {014604} (\bibinfo {year} {2006})}\BibitemShut {NoStop}%
\bibitem [{\citenamefont {Nagahiro}\ \emph {et~al.}(2009)\citenamefont
  {Nagahiro}, \citenamefont {Jido},\ and\ \citenamefont
  {Hirenzaki}}]{Nagahiro:2008rj}%
  \BibitemOpen
  \bibfield  {author} {\bibinfo {author} {\bibfnamefont {H.}~\bibnamefont
  {Nagahiro}}, \bibinfo {author} {\bibfnamefont {D.}~\bibnamefont {Jido}}, \
  and\ \bibinfo {author} {\bibfnamefont {S.}~\bibnamefont {Hirenzaki}},\ }\href
  {\doibase 10.1103/PhysRevC.80.025205} {\bibfield  {journal} {\bibinfo
  {journal} {Phys. Rev.}\ }\textbf {\bibinfo {volume} {C80}},\ \bibinfo {pages}
  {025205} (\bibinfo {year} {2009})}\BibitemShut {NoStop}%
\bibitem [{\citenamefont {Pisarski}\ and\ \citenamefont
  {Wilczek}(1984)}]{Pisarski:1983ms}%
  \BibitemOpen
  \bibfield  {author} {\bibinfo {author} {\bibfnamefont {R.~D.}\ \bibnamefont
  {Pisarski}}\ and\ \bibinfo {author} {\bibfnamefont {F.}~\bibnamefont
  {Wilczek}},\ }\href {\doibase 10.1103/PhysRevD.29.338} {\bibfield  {journal}
  {\bibinfo  {journal} {Phys.Rev.}\ }\textbf {\bibinfo {volume} {D29}},\
  \bibinfo {pages} {338} (\bibinfo {year} {1984})}\BibitemShut {NoStop}%
\bibitem [{\citenamefont {Kunihiro}(1989)}]{Kunihiro:1989my}%
  \BibitemOpen
  \bibfield  {author} {\bibinfo {author} {\bibfnamefont {T.}~\bibnamefont
  {Kunihiro}},\ }\href {\doibase 10.1016/0370-2693(89)90405-X} {\bibfield
  {journal} {\bibinfo  {journal} {Phys. Lett.}\ }\textbf {\bibinfo {volume}
  {B219}},\ \bibinfo {pages} {363} (\bibinfo {year} {1989})}\BibitemShut
  {NoStop}%
\bibitem [{\citenamefont {Kapusta}\ \emph {et~al.}(1996)\citenamefont
  {Kapusta}, \citenamefont {Kharzeev},\ and\ \citenamefont
  {McLerran}}]{Kapusta:1995ww}%
  \BibitemOpen
  \bibfield  {author} {\bibinfo {author} {\bibfnamefont {J.~I.}\ \bibnamefont
  {Kapusta}}, \bibinfo {author} {\bibfnamefont {D.}~\bibnamefont {Kharzeev}}, \
  and\ \bibinfo {author} {\bibfnamefont {L.~D.}\ \bibnamefont {McLerran}},\
  }\href {\doibase 10.1103/PhysRevD.53.5028} {\bibfield  {journal} {\bibinfo
  {journal} {Phys. Rev.}\ }\textbf {\bibinfo {volume} {D53}},\ \bibinfo {pages}
  {5028} (\bibinfo {year} {1996})}\BibitemShut {NoStop}%
\bibitem [{\citenamefont {Costa}\ \emph {et~al.}(2003)\citenamefont {Costa},
  \citenamefont {Ruivo},\ and\ \citenamefont {Kalinovsky}}]{Costa:2002gk}%
  \BibitemOpen
  \bibfield  {author} {\bibinfo {author} {\bibfnamefont {P.}~\bibnamefont
  {Costa}}, \bibinfo {author} {\bibfnamefont {M.~C.}\ \bibnamefont {Ruivo}}, \
  and\ \bibinfo {author} {\bibfnamefont {Y.~L.}\ \bibnamefont {Kalinovsky}},\
  }\href {\doibase 10.1016/S0370-2693(03)00395-2} {\bibfield  {journal}
  {\bibinfo  {journal} {Phys. Lett.}\ }\textbf {\bibinfo {volume} {B560}},\
  \bibinfo {pages} {171} (\bibinfo {year} {2003})}\BibitemShut {NoStop}%
\bibitem [{\citenamefont {Nagahiro}\ \emph {et~al.}(2006)\citenamefont
  {Nagahiro}, \citenamefont {Takizawa},\ and\ \citenamefont
  {Hirenzaki}}]{Nagahiro:2006dr}%
  \BibitemOpen
  \bibfield  {author} {\bibinfo {author} {\bibfnamefont {H.}~\bibnamefont
  {Nagahiro}}, \bibinfo {author} {\bibfnamefont {M.}~\bibnamefont {Takizawa}},
  \ and\ \bibinfo {author} {\bibfnamefont {S.}~\bibnamefont {Hirenzaki}},\
  }\href {\doibase 10.1103/PhysRevC.74.045203} {\bibfield  {journal} {\bibinfo
  {journal} {Phys. Rev.}\ }\textbf {\bibinfo {volume} {C74}},\ \bibinfo {pages}
  {045203} (\bibinfo {year} {2006})}\BibitemShut {NoStop}%
\bibitem [{\citenamefont {Cs\"org\ifmmode~\mbox{\H{o}}\else \H{o}\fi{}}\ \emph
  {et~al.}(2010)\citenamefont {Cs\"org\ifmmode~\mbox{\H{o}}\else \H{o}\fi{}},
  \citenamefont {V\'ertesi},\ and\ \citenamefont
  {Sziklai}}]{PhysRevLett.105.182301}%
  \BibitemOpen
  \bibfield  {author} {\bibinfo {author} {\bibfnamefont {T.}~\bibnamefont
  {Cs\"org\ifmmode~\mbox{\H{o}}\else \H{o}\fi{}}}, \bibinfo {author}
  {\bibfnamefont {R.}~\bibnamefont {V\'ertesi}}, \ and\ \bibinfo {author}
  {\bibfnamefont {J.}~\bibnamefont {Sziklai}},\ }\href {\doibase
  10.1103/PhysRevLett.105.182301} {\bibfield  {journal} {\bibinfo  {journal}
  {Phys. Rev. Lett.}\ }\textbf {\bibinfo {volume} {105}},\ \bibinfo {pages}
  {182301} (\bibinfo {year} {2010})}\BibitemShut {NoStop}%
\bibitem [{\citenamefont {V\'ertesi}\ \emph {et~al.}(2011)\citenamefont
  {V\'ertesi}, \citenamefont {Cs\"org\ifmmode~\mbox{\H{o}}\else \H{o}\fi{}},\
  and\ \citenamefont {Sziklai}}]{PhysRevC.83.054903}%
  \BibitemOpen
  \bibfield  {author} {\bibinfo {author} {\bibfnamefont {R.}~\bibnamefont
  {V\'ertesi}}, \bibinfo {author} {\bibfnamefont {T.}~\bibnamefont
  {Cs\"org\ifmmode~\mbox{\H{o}}\else \H{o}\fi{}}}, \ and\ \bibinfo {author}
  {\bibfnamefont {J.}~\bibnamefont {Sziklai}},\ }\href {\doibase
  10.1103/PhysRevC.83.054903} {\bibfield  {journal} {\bibinfo  {journal} {Phys.
  Rev. C}\ }\textbf {\bibinfo {volume} {83}},\ \bibinfo {pages} {054903}
  (\bibinfo {year} {2011})}\BibitemShut {NoStop}%
\bibitem [{\citenamefont {Moskal}\ \emph
  {et~al.}(2000{\natexlab{a}})\citenamefont {Moskal} \emph
  {et~al.}}]{Moskal:2000gj}%
  \BibitemOpen
  \bibfield  {author} {\bibinfo {author} {\bibfnamefont {P.}~\bibnamefont
  {Moskal}} \emph {et~al.},\ }\href {\doibase 10.1016/S0370-2693(00)00039-3}
  {\bibfield  {journal} {\bibinfo  {journal} {Phys. Lett.}\ }\textbf {\bibinfo
  {volume} {B474}},\ \bibinfo {pages} {416} (\bibinfo {year}
  {2000}{\natexlab{a}})}\BibitemShut {NoStop}%
\bibitem [{\citenamefont {Moskal}\ \emph
  {et~al.}(2000{\natexlab{b}})\citenamefont {Moskal} \emph
  {et~al.}}]{Moskal:2000pu}%
  \BibitemOpen
  \bibfield  {author} {\bibinfo {author} {\bibfnamefont {P.}~\bibnamefont
  {Moskal}} \emph {et~al.},\ }\href {\doibase 10.1016/S0370-2693(00)00533-5}
  {\bibfield  {journal} {\bibinfo  {journal} {Phys. Lett.}\ }\textbf {\bibinfo
  {volume} {B482}},\ \bibinfo {pages} {356} (\bibinfo {year}
  {2000}{\natexlab{b}})}\BibitemShut {NoStop}%
\bibitem [{\citenamefont {Nanova}(2010)}]{Nanova:2011baryons}%
  \BibitemOpen
  \bibfield  {author} {\bibinfo {author} {\bibfnamefont {M.}~\bibnamefont
  {Nanova}},\ }\href@noop {} {} (\bibinfo {year} {2010}),\ \bibinfo {note}
  {talk in Baryons10, Dec 7-12, 2010, Osaka Univ., Osaka, Japan.}\BibitemShut
  {Stop}%
\bibitem [{\citenamefont {Veneziano}(1989)}]{Veneziano:1989ei}%
  \BibitemOpen
  \bibfield  {author} {\bibinfo {author} {\bibfnamefont {G.}~\bibnamefont
  {Veneziano}},\ }\href {\doibase 10.1142/S0217732389001830} {\bibfield
  {journal} {\bibinfo  {journal} {Mod. Phys. Lett.}\ }\textbf {\bibinfo
  {volume} {A4}},\ \bibinfo {pages} {1605} (\bibinfo {year}
  {1989})}\BibitemShut {NoStop}%
\bibitem [{\citenamefont {Veneziano}(1979)}]{Veneziano:1979ec}%
  \BibitemOpen
  \bibfield  {author} {\bibinfo {author} {\bibfnamefont {G.}~\bibnamefont
  {Veneziano}},\ }\href {\doibase 10.1016/0550-3213(79)90332-8} {\bibfield
  {journal} {\bibinfo  {journal} {Nucl. Phys.}\ }\textbf {\bibinfo {volume}
  {B159}},\ \bibinfo {pages} {213} (\bibinfo {year} {1979})}\BibitemShut
  {NoStop}%
\bibitem [{\citenamefont {Christos}(1984)}]{Christos:1984tu}%
  \BibitemOpen
  \bibfield  {author} {\bibinfo {author} {\bibfnamefont {G.~A.}\ \bibnamefont
  {Christos}},\ }\href {\doibase 10.1016/0370-1573(84)90025-5} {\bibfield
  {journal} {\bibinfo  {journal} {Phys. Rept.}\ }\textbf {\bibinfo {volume}
  {116}},\ \bibinfo {pages} {251} (\bibinfo {year} {1984})}\BibitemShut
  {NoStop}%
\bibitem [{\citenamefont {Inoue}\ and\ \citenamefont
  {Oset}(2002)}]{Inoue:2002xw}%
  \BibitemOpen
  \bibfield  {author} {\bibinfo {author} {\bibfnamefont {T.}~\bibnamefont
  {Inoue}}\ and\ \bibinfo {author} {\bibfnamefont {E.}~\bibnamefont {Oset}},\
  }\href@noop {} {\bibfield  {journal} {\bibinfo  {journal} {Nucl. Phys.}\
  }\textbf {\bibinfo {volume} {A710}},\ \bibinfo {pages} {354} (\bibinfo {year}
  {2002})}\BibitemShut {NoStop}%
\bibitem [{\citenamefont {Mares}\ \emph {et~al.}(2006)\citenamefont {Mares},
  \citenamefont {Friedman},\ and\ \citenamefont {Gal}}]{Mares:2006vk}%
  \BibitemOpen
  \bibfield  {author} {\bibinfo {author} {\bibfnamefont {J.}~\bibnamefont
  {Mares}}, \bibinfo {author} {\bibfnamefont {E.}~\bibnamefont {Friedman}}, \
  and\ \bibinfo {author} {\bibfnamefont {A.}~\bibnamefont {Gal}},\ }\href
  {\doibase 10.1016/j.nuclphysa.2006.02.010} {\bibfield  {journal} {\bibinfo
  {journal} {Nucl. Phys.}\ }\textbf {\bibinfo {volume} {A770}},\ \bibinfo
  {pages} {84} (\bibinfo {year} {2006})}\BibitemShut {NoStop}%
\bibitem [{\citenamefont {Jido}\ \emph {et~al.}(2002)\citenamefont {Jido},
  \citenamefont {Nagahiro},\ and\ \citenamefont {Hirenzaki}}]{Jido:2002yb}%
  \BibitemOpen
  \bibfield  {author} {\bibinfo {author} {\bibfnamefont {D.}~\bibnamefont
  {Jido}}, \bibinfo {author} {\bibfnamefont {H.}~\bibnamefont {Nagahiro}}, \
  and\ \bibinfo {author} {\bibfnamefont {S.}~\bibnamefont {Hirenzaki}},\
  }\href@noop {} {\bibfield  {journal} {\bibinfo  {journal} {Phys. Rev.}\
  }\textbf {\bibinfo {volume} {C66}},\ \bibinfo {pages} {045202} (\bibinfo
  {year} {2002})}\BibitemShut {NoStop}%
\bibitem [{\citenamefont {Nagahiro}\ \emph {et~al.}(2003)\citenamefont
  {Nagahiro}, \citenamefont {Jido},\ and\ \citenamefont
  {Hirenzaki}}]{Nagahiro:2003iv}%
  \BibitemOpen
  \bibfield  {author} {\bibinfo {author} {\bibfnamefont {H.}~\bibnamefont
  {Nagahiro}}, \bibinfo {author} {\bibfnamefont {D.}~\bibnamefont {Jido}}, \
  and\ \bibinfo {author} {\bibfnamefont {S.}~\bibnamefont {Hirenzaki}},\ }\href
  {\doibase 10.1103/PhysRevC.68.035205} {\bibfield  {journal} {\bibinfo
  {journal} {Phys. Rev.}\ }\textbf {\bibinfo {volume} {C68}},\ \bibinfo {pages}
  {035205} (\bibinfo {year} {2003})}\BibitemShut {NoStop}%
\bibitem [{\citenamefont {Nagahiro}\ \emph {et~al.}(2005)\citenamefont
  {Nagahiro}, \citenamefont {Jido},\ and\ \citenamefont
  {Hirenzaki}}]{Nagahiro:2005gf}%
  \BibitemOpen
  \bibfield  {author} {\bibinfo {author} {\bibfnamefont {H.}~\bibnamefont
  {Nagahiro}}, \bibinfo {author} {\bibfnamefont {D.}~\bibnamefont {Jido}}, \
  and\ \bibinfo {author} {\bibfnamefont {S.}~\bibnamefont {Hirenzaki}},\ }\href
  {\doibase 10.1016/j.nuclphysa.2005.07.001} {\bibfield  {journal} {\bibinfo
  {journal} {Nucl. Phys.}\ }\textbf {\bibinfo {volume} {A761}},\ \bibinfo
  {pages} {92} (\bibinfo {year} {2005})}\BibitemShut {NoStop}%
\bibitem [{\citenamefont {Jido}\ \emph
  {et~al.}(2008{\natexlab{b}})\citenamefont {Jido}, \citenamefont
  {Kolomeitsev}, \citenamefont {Nagahiro},\ and\ \citenamefont
  {Hirenzaki}}]{Jido:2008ng}%
  \BibitemOpen
  \bibfield  {author} {\bibinfo {author} {\bibfnamefont {D.}~\bibnamefont
  {Jido}}, \bibinfo {author} {\bibfnamefont {E.~E.}\ \bibnamefont
  {Kolomeitsev}}, \bibinfo {author} {\bibfnamefont {H.}~\bibnamefont
  {Nagahiro}}, \ and\ \bibinfo {author} {\bibfnamefont {S.}~\bibnamefont
  {Hirenzaki}},\ }\href {\doibase 10.1016/j.nuclphysa.2008.07.012} {\bibfield
  {journal} {\bibinfo  {journal} {Nucl. Phys.}\ }\textbf {\bibinfo {volume}
  {A811}},\ \bibinfo {pages} {158} (\bibinfo {year}
  {2008}{\natexlab{b}})}\BibitemShut {NoStop}%
\bibitem [{\citenamefont {Rader}\ \emph {et~al.}(1972)\citenamefont {Rader}
  \emph {et~al.}}]{Rader:1973mx}%
  \BibitemOpen
  \bibfield  {author} {\bibinfo {author} {\bibfnamefont {R.}~\bibnamefont
  {Rader}} \emph {et~al.},\ }\href {\doibase 10.1103/PhysRevD.6.3059}
  {\bibfield  {journal} {\bibinfo  {journal} {Phys.Rev.}\ }\textbf {\bibinfo
  {volume} {D6}},\ \bibinfo {pages} {3059} (\bibinfo {year}
  {1972})}\BibitemShut {NoStop}%
\bibitem [{\citenamefont {Nagahiro}(2010)}]{Nagahiro:2010zz}%
  \BibitemOpen
  \bibfield  {author} {\bibinfo {author} {\bibfnamefont {H.}~\bibnamefont
  {Nagahiro}},\ }\href {\doibase 10.1143/PTPS.186.316} {\bibfield  {journal}
  {\bibinfo  {journal} {Prog. Theor. Phys. Suppl.}\ }\textbf {\bibinfo {volume}
  {186}},\ \bibinfo {pages} {316} (\bibinfo {year} {2010})}\BibitemShut
  {NoStop}%
\end{thebibliography}
\end{document}